**Optimization of salt concentration and Explanation of Two Peak Percolation in Blend Solid Polymer Nanocomposite Films**

Anil Arya, A. L. Sharma*

*Centre for Physical Sciences, Central University of Punjab, Bathinda, Punjab-151001, INDIA*

E-mail: alsharma@cup.edu.in

**Abstract**

The present paper report is focused toward the preparation of the flexible and freestanding blend solid polymer electrolyte films based on PEO-PVP complexed with $NaPF_6$ by solution cast technique. The structural/morphological features of the synthesized polymer nanocomposite films have been investigated in detail using X-ray diffraction, Fourier transform infra-red spectroscopy, field emission scanning electron microscope, and atomic force microscopy techniques. The film PEO-PVP+$NaPF_6$ ($Ö/Na^+ = 8$) exhibits highest ionic conductivity ~$5.92 \times 10^{-6}$ S cm$^{-1}$ at 40 °C and ~$2.46 \times 10^{-4}$ S cm$^{-1}$ at 100 °C. The temperature dependent conductivity shows Arrhenius type behavior and activation energy decreases with the addition of salt. The high temperature (100 °C) conductivity monitoring is done for the optimized PEO-PVP+$NaPF_6$ ($Ö/Na^+ = 8$) highly conductive system and the conductivity is still maintained stable up to 160 h (approx. 7 days). The thermal transitions parameters were measured by the differential scanning calorimetry (DSC) measurements. The prepared polymer electrolyte film displays the smoother surface in addition of salt and a thermal stability up to 300 ºC. The ion transference number ($t_{ion}$) for the highest conducting sample is found to be 0.997 and evidence that the present system is ion dominating with negligible electron contribution. Both linear sweep voltammetry and cyclic voltammetry supports the use of prepared polymer electrolyte with long-term cycle stability and thermal stability for the solid state sodium ion batteries. Finally, a two peak percolation mechanism has been proposed on the basis of experimental findings.

**Introduction**

The most challenging and discussible issue is the search of the sustainable & appropriate source of energy that can fulfill the increasing global demand. The use of renewable and cleaner energy sources, such as solar radiation, wind, and waves, is flattering as a key source of energy which can amend the dependency on the non-renewable sources of energy (fossil fuels) and will lower the global temperature by 2 ºC. The most attractive and suitable candidate for the clean & efficient energy source is secondary batteries such as a lithium-ion battery (LIB) and sodium ion battery (SIB). The former one is main thrust area of research and mostly used in the cellular phone, personal computers, electric vehicles, hybrid electric vehicle and other digital portable products, due to high specific energy and power. LIB has associated issues such as safety, lack of abundance and high cost. So, to develop an alternative abundant, cheap, stable and non-toxic energy storage system has pushed forward R&D on SIB and demonstrating itself as a potentially more convenient alternative to existing LIB [1-3]. Due to the increasing global demand for energy, SIB research boosted itself due to the abundance (6[th] most abundant element in earth crust, 2.64 %), readily accessible in both earth-crust/ocean, high reduction potential (-2.7 V), low toxicity, low cost (seven times lower than lithium, softness and atomic mass). To meet the global demand for energy, now research is focused on the development of sodium ion based devices due to the uniform geographical distribution of sodium [4-12].

A conventional battery is comprised of a cathode, anode, separator, and electrolyte. Both cathode and anode are the accommodating sites for the ions. The separator role is to physically separate the electrodes while electrolyte provides the medium to the ions shuttling between electrodes. At present, the liquid electrolyte is used in commercial systems and that limit its geometry and application range due to an associated drawback such as leakage of organic solvents, flammability, short-circuit issue, poor mechanical properties, and incompatibility with high energy batteries. So, there is need of suitable electrolyte cum separator with desirable ionic conductivity and the long cycle stability enabling better performance. The polymer electrolytes are alternate to the liquid electrolyte and are attempted since the first report by Fenton et al., [13] in 1973. Then in 1978 polymer electrolytes were proposed first time for application in batteries due to advantageous of both solid state electrochemistry and easy preparation [14]. Then gel polymer electrolyte (GPE) prepared by incorporation of the plasticizer emerged as an attractive candidate due to good compatibility with liquid electrolyte but poor mechanical property, interfacial properties and deficiency of shutdown behavior was a bigger constraint for commercial applications [15]. Solid polymer electrolytes (SPEs) are the potential candidate to overcome the above-said drawback associated with the traditional liquid and gel polymer electrolyte. Soin order to the replacement of liquid and gel polymer electrolyte completely, SPE must possess desirable ionic conductivity, enhanced thermal, electrochemical and mechanical stability. One major advantage with the SPE is the simple & low-cost design strategy, flexibility, and miniaturization of devices that automatically lowers both cost and weight. As no liquid part is used, so all solid state battery guarantees great safety than the existing one [16-17]. SPE also plays a dual role in energy storage/conversion devices. It permits the transportation of ionic charge carriers as well as prevent electrical short circuits between the electrodes. So, various polymer materials have been studied and developed as SPE matrix [18-23].

Poly(ethylene oxide) is one of the most auspicious candidates since last three decades due to its low glass transition temperature, the good capability to dissolve salts, high ionic conductivity, and high degradation temperature. The presence of electron donor ether group $(CH_2 - \ddot{O} - CH_2)$ in polymer backbone $(-CH_2 - CH_2 - \ddot{O} -)$ makes it fascinating for coordination with the available cations and favor enhanced polymer segmental motion. However, the semi-crystalline nature of PEO is responsible for low ionic conductivity, and poor mechanical strength hinders itS usage in SPEs. Various strategies have been adopted to improve the inclusive performance of PEO-based electrolytes by modifying the structure of the PEO. Since it is well proven that enormous amorphous phase favors easier and smoother cation migration [24-27] in such ion conducting system. Therefore, polymer blending approach is tried to develop to make such system more amorphous and appears appropriate in improving the compromised parameters during use of a semi-crystalline polymer as host materials. Polymer blends are obtained by physical mixing of two or more polymers without any chemical reaction among them. One sole advantage of blending approach is appropriate to control over the properties by varying the material composition and easiest means of preparation. Recently, many efforts have been devoted in enhancing the electrochemical and mechanical properties in blend solid polymers electrolytes (BSPE) membranes such as PEO-PAN, PVdF-PEO, PEO-PVdF, PVA-PEO, PEO-PEG, PVA-PVP, PVC-PEMA, PVC-PVdF, PEO-PVP, PEO-P(VdF-HFP) and PVC-PEO [28-42].

Poly(vinyl pyrrolidone) (PVP) has some remarkable properties that motivated us for its selection as the partner with PEO (semi-crystalline polymer) in the preparation of blend polymer electrolyte (BPE). It possesses moderate electrical conductivity, good environmental stability, rich physics in charge transport mechanism and easiest way of preparation. Also, the high amorphous content associated with it validates its candidature. The presence of the rigid pyrrolidone group in amorphous PVP helps to provide better ionic mobility in these systems and carbonyl group (C=O) attached to the side chains of PVP helps in the formation of a number of complexes with different inorganic salts [43-46].

Doping of sodium salt in existing technology (lithium based electrolyte) overcomes the other issues like smaller cations such as lithium capturing probability by the polymer network are more and that lowers the ion mobility [47]. Another advantage of the sodium is that its softness nature makes the smooth and proper contact with the electrochemical devices. The ionic conducting properties of PEO, doped with different sodium salts like sodium perchlorate ($NaClO_4$), sodium fluoride (NaF), sodium iodide (NaI), sodium bromide (NaBr), sodium hexafluoro phosphate ($NaPF_6$), sodium periodate ($NaIO_4$), sodium tetrafluoroborate ($NaBF_4$) have already been reported in literature [10, 46, 48-55]. So, after reviewing the various sodium based solid polymer electrolyte systems, sodium hexafluoro phosphate ($NaPF_6$) is selected as the cation ($Na^+$) source for doping PEO/PVP based blend solid polymer electrolyte. The current report is focused toward the search of suitable SPE by the complexation of $NaPF_6$ with blend PEO-PVP. So, here the advantage of flexible high solvation matrix of PEO and amorphous content of PVP for sodium ions are utilized for developing the blend solid polymer electrolyte [56-59] system.

As per author finding, there are not any studies based on $NaPF_6$ incorporated PEO/PVP based blend polymer electrolyte for the development of flexible and freestanding SPE films. So, in the present investigation, the optimum concentration of $NaPF_6$ in PEO/PVP blend polymer is obtained to develop good quality free standing SPE films. First, the structural, surface and microstructural investigations are established by X-ray diffraction (XRD), field emission scanning electron microscopy (FE-SEM), atomic force microscopy (AFM) and Fourier transformation infra-Red spectroscopy (FTIR). Thermo-gravimetric analysis (TGA) is used to find the thermal stability of the synthesized film sample and differential scanning calorimetry (DSC) to find the glass transition temperature and crystallinity of all SPEs. The ionic conductivity of the prepared solid polymer electrolyte samples at different temperatures is measured using an electrochemical analyzer set up. The ion transference number of the electrolyte film is estimated using dc polarization method. The linear sweep voltammetry (LSV) and cyclic voltammetry (CV) were characterized for obtaining the voltage window of the electrolyte and also the temperature dependency of the voltage window was observed.

To get a better insight on the cation transport within the electrolytes, the correlation between the estimated transport parameters with electrical parameters has also been analyzed. Finally, a mechanism is proposed that supports the experimental findings.

**Experimental**

*Materials*

PEO ($M_w$=200,000), PVP ($M_w$=40000) and $NaPF_6$ ($M_w$=167.95 g/mol) were purchased from Sigma-Aldrich, India and used as received. The chemical structure of all material is displayed in figure 1. The solvent methanol was purchased from Lobha Chemicals, India. The standard solution cast technique was used for the preparation of PEO-PVP+NaPF$_6$ complex with methanol as common solvent.

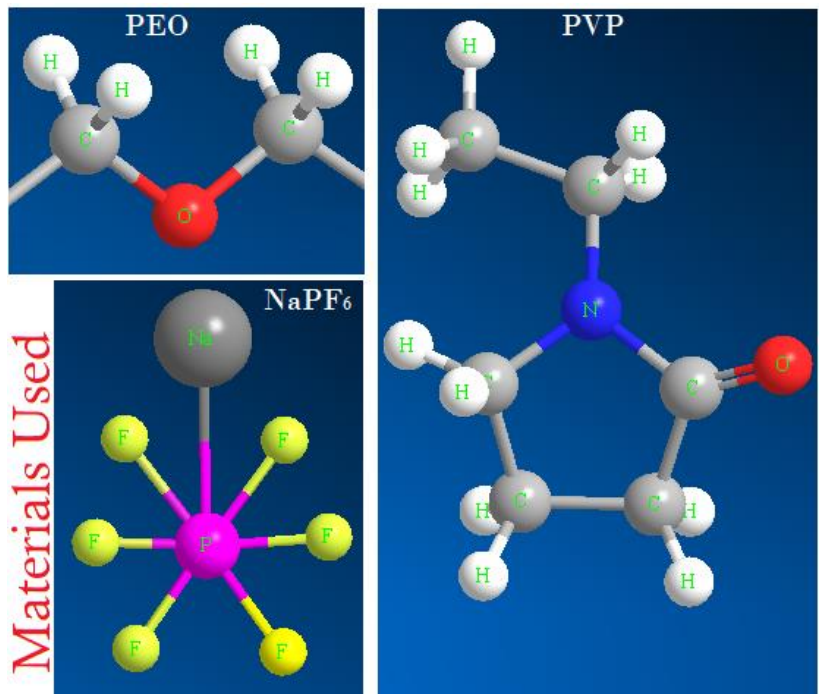

Figure 1. Structure unit of PEO, PVP and salt NaPF$_6$.

*Preparation of PEO-PVP+NaPF$_6$ based solid polymer electrolyte*

The concentration of PEO and PVP were kept constant 80:20 for all solid polymer electrolytes. The addition of an appropriate amount of salt (NaPF$_6$) stoichiometric ratio has been calculated considering oxygen of PEO. The formula for this calculation is shown below;

For PEO:

$$\frac{\text{Ö}}{\text{Na}^+} = \frac{\text{No. of monomer unit in 0.4 gram of PEO}}{\text{No. of NaPF}_6 \text{ molecule in half gram of salt}} \times \frac{\text{wt. of PEO took}}{\text{wt. of salt is taken}}$$

The salt concentration (Ö/Na$^+$) was varied as Ö/Na$^+$ = 2, 4, 6, 8, 10.

First of all PEO and PVP were added in methanol (20 ml) and kept for 10 min for swelling of polymer chains. Then stirring was done for 4 h at room temperature, until a completely transparent and homogeneous solution was obtained. Then the appropriate stoichiometric ratio of salt (Ö/Na$^+$ = 2, 4, 6, 8, 10) was added and stirring was done again for 12 h at room temperature. The final homogenous solution obtained was cast in the Petri-dishes and kept at room temperature in a desiccator (silica gel inside) to avoid form any moisture content. Then after evaporation of the solvent, for complete removal of solvent petri-dish were kept in a vacuum oven at 60 °C for 24 h. Finally, the free-standing solid polymer electrolyte films were obtained after peeling off from the petri dish. The prepared film was stored in a desiccator to avoid contamination and for further relevant characterizations. The blend formation interaction mechanism (Figure 2a) along with a snapshot of solution cast technique is depicted in Figure 2b.

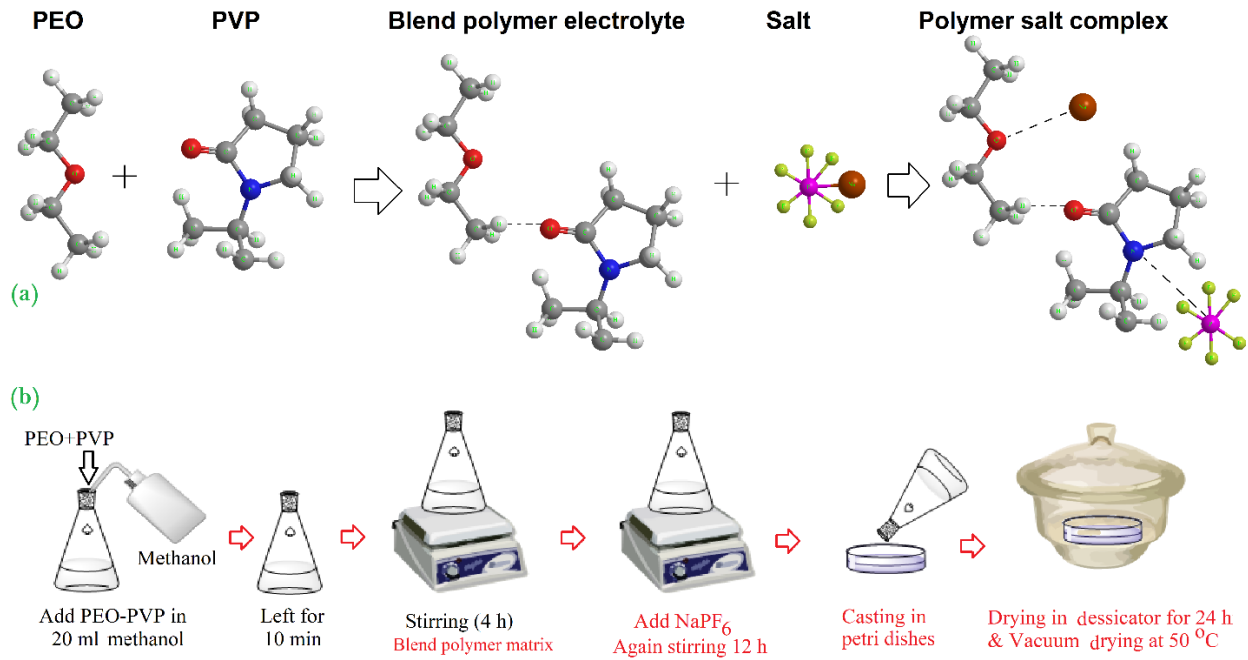

Figure 2. **(a)** Representation of blend formation and cation coordination, **(b)** flow chart of solution cast technique.

*Characterization*

X-ray diffraction (XRD) (Bruker D8 Advance) was performed for the determination of crystallinity and recorded with Cu-K$_\alpha$ radiation ($\lambda$ = 1.54 Å) in the Braggs angle range ($2\theta$) from 10° to 60°. Field emission scanning electron microscopy (FESEM) was used to study the surface morphology (FESEM: Carl Zeiss product) and taken in a high vacuum after sputtering the samples with gold in order to prepare conductive surfaces. The Fourier transform infrared (FTIR) spectra (Bruker Tensor 27, Model: NEXUS–870) were recorded in absorbance mode over the wavenumber region from 600 to 3500 cm$^{-1}$ (resolution of 4 cm$^{-1}$) to probe the presence of various microscopic interactions such as polymer-ion, ion-ion interaction and complex formation. The ionic conductivity was measured by impedance spectroscopy (IS) in the frequency range of 1 Hz to 1 MHz using the CHI 760 electrochemical analyzer. An AC sinusoidal signal of 10 mV was applied to the cell configuration SS|SPE|SS where solid polymer electrolytes films were sandwiched between two stainless steel (SS) electrodes. The intercept between the semi-circle at high frequency and tilted spike at low frequency were taken as the bulk resistance ($R_b$). The electrical conductivity ($\sigma$) value was obtained using equation 1:

$$\sigma_{dc} = \frac{1}{R_b}\frac{t}{A} \quad (1)$$

Where 't' is thickness (cm) of the polymer film (100-125 μm), $R_b$ is bulk resistance ($\Omega$) and A is area (cm$^2$) of working electrode (1.43 cm$^2$). The variation of ionic conductivity of the electrolyte film with temperature was studied in the temperature range from 40 °C to 100 °C with a temperature difference of 10 °C (Temperature Controller; Marine India). The thermal activation energy for ionic transport was estimated from the slope of the linear fit of the Arrhenius plot. The linear variation in log ($\sigma$/S cm$^{-1}$) vs. 1000/T plot suggests a thermally activated process represented by $\sigma =$ $\sigma_o \exp(-E_a/kT)$, where $\sigma_o$ is the constant pre-exponential factor and $E_a$ is the activation energy. The parameter $T$ stands for the absolute temperature and $k$ for the Boltzmann constant.

Differential scanning calorimetry (DSC) measurements were performed to find the glass transition temperature, melting temperature and crystallinity of all SPEs with a heating rate of 10 °C min$^{-1}$ from -100 to 100 ºC under an N$_2$/Ar atmosphere (DSC-Sirius 3500). SPEs films with the weight of 8-10 mg were sealed in aluminum pans, and an empty sealed aluminum pan was used as a reference. The total ionic transference number ($t_{ion}$) was obtained by placing polymer electrolyte film between stainless steel (SS) blocking electrodes and a fixed dc volatge of 10 mV was applied across the *SS/SPE/SS* cell. Ion transference numbers of the solid polymer electrolytes was evaluated using equation 2 on *SS/SPE/SS* cell:

$$t_{ion} = \left(\frac{I_t - I_e}{I_t}\right) \times 100 \qquad (2)$$

The prepared polymer electrolytes have also been subjected to atomic force microscopy (AFM, Veeco CP-II) surface image studies. Thermal stability of the synthesized SPE films was investigated using thermo-gravimetric analysis (TGA – SHIMADZU–DTG-60H) under dynamic temperature conditions from 30 °C to 600 °C, in a controlled nitrogen atmosphere at a constant heating scan rate of 10 °C min$^{-1}$. The linear sweep voltammetry (LSV) and cyclic voltammetry (CV) were characterized using CHI 760 electrochemical analyzer for obtaining the voltage window of the electrolyte. The sample codes are SPE1, SPE2, SPE3, SPE4, SPE 5 and SPE6 for blend solid polymer electrolyte films with Ö/Na$^+$ = 0, 2, 4, 6, 8, 10 of NaPF$_6$ respectively.

**Results and Discussions**

**X-ray diffraction analysis**

XRD is an important tool to investigate about the complex formation and change in peak position, peak intensity provides necessary information that helps us to understand the role played by salt in the blend polymer electrolyte (BPE). Figure 3 depicts the XRD diffractograms of PEO-PVP blend with different stoichiometric salt content (Ö/Na$^+$ = 2, 4, 6, 8, 10) in the 2 theta range 10º to 60º. NaPF$_6$ exhibits sharp and intense diffraction peaks near 20º, 22º, 32º, 44º along with two minor peaks near 32º. The absence of any sharp salt associated peak indicates the complete dissociation of the salt. The absence of any peak associated with salt suggests the complete dissociation of the salt. All SPE shows an almost identical pattern with two crystalline peaks of PEO at 19º and 23º with some additional peaks on the addition of salt which indicates the complex formation. The additional peaks that are not of pure PEO and pure NaPF$_6$ may be due to some sort of a long-range order set by the presence of ion multiplets (as Na$_2$X$^+$, NaX$_2^-$, etc.) [48, 60-61]. These two peaks at 19º and 23º are associated with the (120) and (112/032) plane, respectively. The shift in peak position confirms that the polymer blend structure is modified after addition of salt (Ö/Na$^+$ = 2, 4, 6, 8, 10) and it evidences that salt plays an effective role in altering the structure of blend polymer matrix. The reduction of peak intensity (at 19º and 23º) indicates the lowering of the crystallinity which evidences the enhancement of amorphous content owing to the cation coordination with the electron rich group of PEO/PVP. The small peaks at 13º and a broad peak near 21º are associated with the amorphous nature of PVP. The absence of any sharp peak on high salt content evidences the presence of amorphous content. This results in enhancement of ionic conductivity may be due to improved flexibility associated with the amorphous phase and faster segmental motion of the polymer chains.

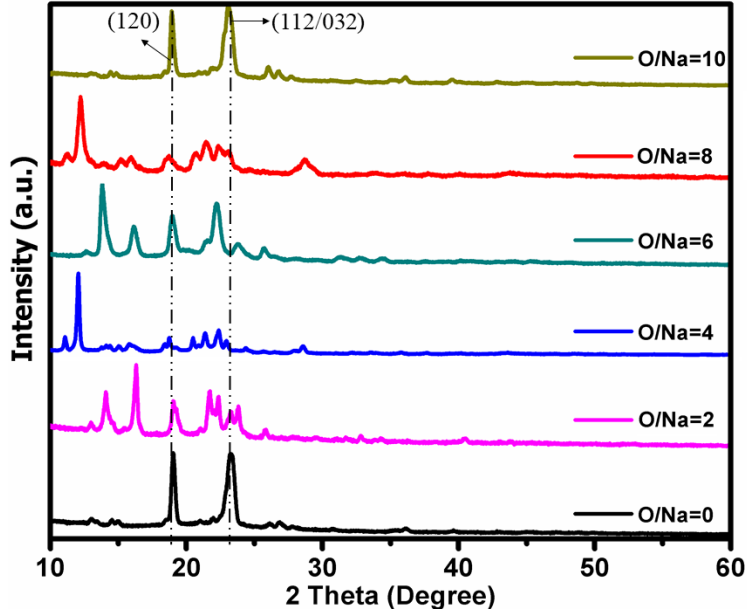

**Figure 3.** XRD diffractograms for PEO-PVP+NaPF$_6$ (Ö/Na$^+$ = 2, 4, 6, 8, 10)

For, Ö/Na$^+$ = 8 a broad hump with decreased intensity as compared to Ö/Na$^+$ = 10 is observed between the 12º to 25º and may be due to the interaction between the cation and salt. Thus, with the addition of salt Ö/Na$^+$ = 4 some new peaks are generated which may be due to the incomplete dissociation of the salt in the blend polymer electrolyte matrix. Few, minor peaks near 21º may be attributed to the insufficient interaction between the polymer matrix and salt that leads to ion association. Also, some peak shows splitting that is attributed to the semi-crystalline (partially crystalline-partly amorphous) nature of the solid polymer electrolyte. This needs to be validated further by more evidence as discussed in the next section.

When different salt content is added to the polymer blend then the peaks at 19º and a 23º shift toward the lower angle side. This indicates the increase of interlayer spacing (*d*-spacing) and the interchain separation (*R*), which reveals the complexation on the addition of salt and enhancement in amorphous content as shown in Table 1. The d-spacing between the diffraction planes was obtained using the Bragg's formula $2d\sin\theta=n\lambda$ and interchain separation (*R*) using the equation $R=5\lambda/8\sin\theta$ [62]. As anion is going to be coordinated with the polymer backbone while cation with electron rich group this overall disrupts the ordering of the chains and this directly supports the enhancement of the amorphous content. This lowers the covalent bonding in between the polymer chains and cation get more free volume for migration. The highest in both interlayer spacing (*d*-spacing) and the interchain separation (*R*) was for the SPE 5 (Ö/Na$^+$ = 8) and this suggests that the electric properties will be superior to this salt content due to high structure disorder.

Table 1. Values of *2θ* (degree), *d*-spacing (Å) and *R* (Å) of PEO-PVP+NaPF$_6$ (Ö/Na$^+$ = 2, 4, 6, 8, 10) for (120) diffraction peak.

| Sample | $2\theta$ (degree) | $d$-spacing (Å) | $R$ (Å) |
|--------|-------------------|-----------------|---------|
| SPE 1  | 19.03             | 4.66            | 5.82    |
| SPE 2  | 19.03             | 4.66            | 5.82    |
| SPE 3  | 18.78             | 4.71            | 5.89    |
| SPE 4  | 18.92             | 4.68            | 5.85    |
| SPE 5  | 18.70             | 4.73            | 5.92    |
| SPE 6  | 18.91             | 4.68            | 5.85    |

**FESEM analysis**

The ionic transport in the solid polymer electrolyte is associated with the homogeneity of the electrolyte film, FESEM is performed to get more insights of the role played by salt in blend polymer matrix.

Figure 4 depicts the FESEM micrographs of the pure PEO, PEO-PVP blend, SPE 5 with composition [PEO-PVP+NaPF$_6$ (Ö/Na$^+$ =8)] and elemental mapping of Na, P, F in the SPE 5 system. The micrograph of pure PEO shows a rough surface with micro-cracks which is characteristic nature of pure PEO (Figure 4 a) and is attributed to the semi-crystalline nature of the PEO. Addition of PVP in the pure PEO leads to blending and surface modification occurs as shown in Figure 4 b. This surface modification confirms the suitable interactions between both polymers. This change in the surface morphology is an indication of the reduction of crystallinity, and hence, the enhanced amorphous content. As smooth and homogenous surface morphology is evidence of fast ion transport in solid state ionic conductor. Further, the addition of salt in the polymer blend smoothens the surface which reveals the disruption of the crystalline phase (Figure 4 c). This enhancement in the amorphous content of the polymer salt complex will facilitate the fast ion transport. Another remarkable point to be noted is that the smoothening of the surface also increases the solid polymer electrolyte film flexibility which is desirable in the field of solid state ionic conductors. Inset of Figure 4c displays the EDX spectra of Na, P, F atoms which confirm their presence in the polymer salt matrix. Furthermore, the elemental mapping is implemented for imagining the Na, P and F atoms (Figure 4 d). The uniform distribution of Na, P, F in the entire micrograph suggests the complete dissociation of the salt, hence the complex formation is evidenced. The smoothened surface morphology may be reflected in the enhanced electrochemical properties as explained in the upcoming section.

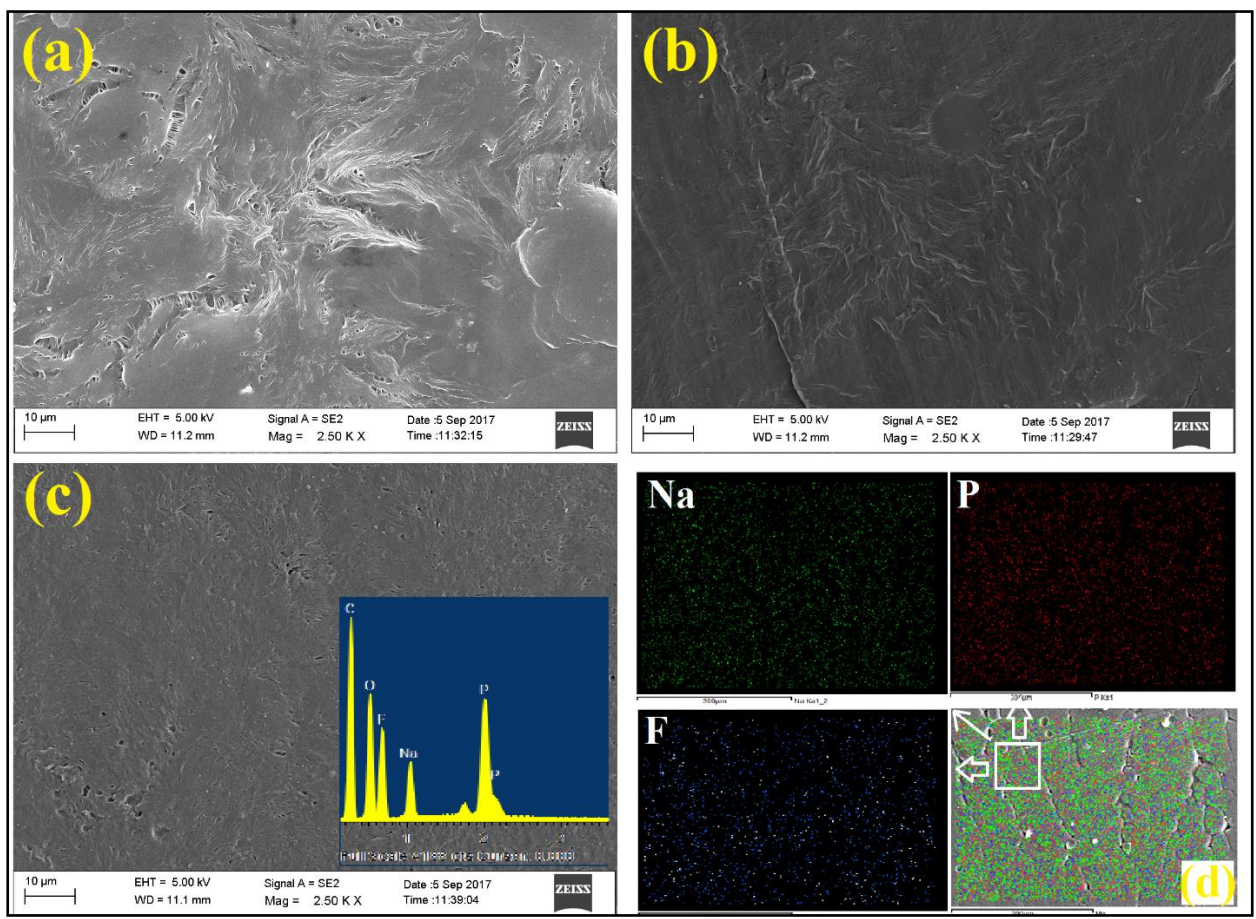

Figure 4. FESEM micrographs of **(a)** pure PEO, **(b)** PEO-PVP blend, **(c)** SPE 5; PEO-PVP+NaPF$_6$ (Ö/Na$^+$ =8) [Inset in shows EDX spectra] and **(d)** elemental mapping of Na (green dot), P (red dot), F (blue dot) for SPE 5.

**FTIR analysis**

FTIR spectral analysis is a powerful tool to explore the molecular interaction, modification of chemical bonding in polymer electrolyte on the addition of salt [63]. Figure 5 depicts the FTIR absorption bands of PEO-PVP and PEO-PVP+NaPF$_6$ (Ö/Na$^+$ = 2, 4, 6, 8, 10) in the wavenumber region 600-3000 cm$^{-1}$ and fingerprint region is shown by dotted line.

The characteristics vibrations bands of PEO and PVP are noticed clearly in the PEO-PVP blend. The C-O stretching vibration located at 953 cm$^{-1}$ is associated with PEO with some CH$_2$ rocking asymmetric vibration. The band regarded at 1117 cm$^{-1}$ is attributed to the asymmetric C-O-C stretching and is a characteristic band of PEO. The vibration bands observed at 1282 cm$^{-1}$, 1348 cm$^{-1}$, 1461 cm$^{-1}$ are assigned to CH$_2$ asymmetric twisting, CH$_2$ bending of PEO and CH$_2$ wagging, respectively. Two fundamental characteristic bands of PEO at 2882 cm$^{-1}$ and near 2900 cm$^{-1}$ correspond to the symmetric C-H and asymmetric C-H stretching respectively [57]. The band located at 845 cm$^{-1}$ is owing to the CH$_2$ rocking mode of PVP with little bit contribution from C–O stretching mode of host polymer PEO. The C–C bending mode of vibration appears on the addition of the salt in polymer blend at 1078 cm$^{-1}$. The absorption band located near 1461 cm$^{-1}$ belongs to the CH$_2$ wagging of PVP. Furthermore, PVP presence is verified by the location of two strong absorption bands at 1348 cm$^{-1}$ and 1687 cm$^{-1}$ which corresponds to the C-N stretching and C=O stretching, respectively [56, 64].

Table 2. FTIR spectral data of PEO-PVP, PEO-PVP+NaPF$_6$ based SPE films.

| Wavenumber (cm$^{-1}$) | | | | | | Band assignment |
|---|---|---|---|---|---|---|
| SPE 1 | SPE 2 | SPE 3 | SPE 4 | SPE 5 | SPE 6 | |
| 844 | 832 | 842 | 844 | 842 | 844 | C-O stretching in PEO/PF$_6^-$ vibration and CH$_2$ rocking of PVP |
| 953 | 955 | 950 | 938 | 943 | 953 | C-O stretching vibration |
| 1117 | 1078/1120 | 1080/1120 | 1080/1127 | 1083/1130 | 1090/1132 | Symmetric and asymmetric C-O-C stretching (amorphous) |
| 1282 | 1245/1287 | 1242/1289 | 1248/1287 | 1242/1289 | 1279 | CH$_2$ symmetric/ asymmetric twisting |
| 1348 | 1350 | 1353 | 1360 | 1355 | 1350 | CH$_2$ bending of PEO and C-N stretching of PVP |
| 1461 | 1468 | 1466 | 1459 | 1459 | 1460 | CH$_2$ wagging |
| 1657 | 1643 | 1655 | 1660 | 1662 | 1658 | C=O stretching |
| 2882 | 2891 | 2844 | 2886 | 2880 | 2861 | Symmetric C-H stretching |
| - | 2948 | 2945 | 2938 | 2932 | 2911 | Asymmetric C-H stretching/CH$_2$ asymmetrical vibration |

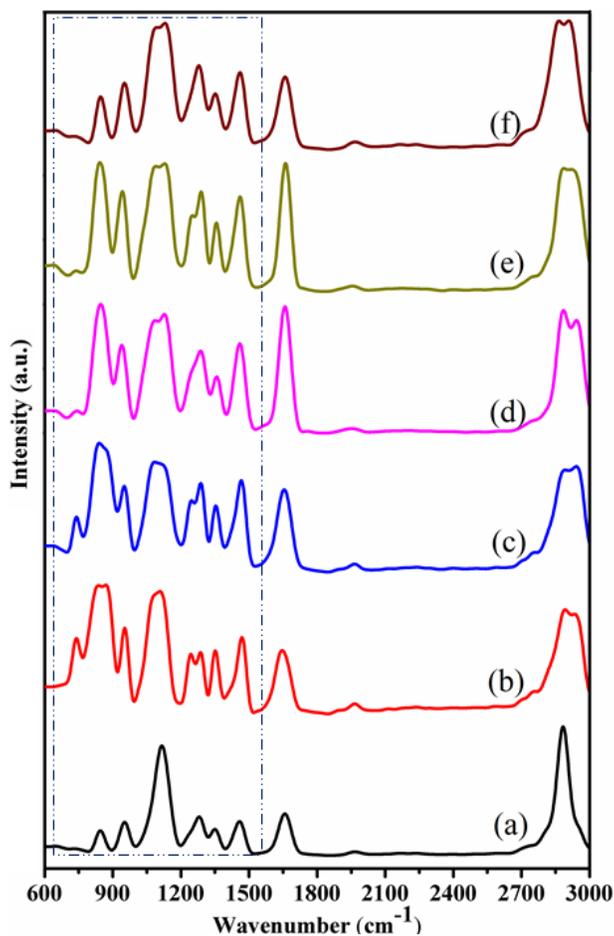

Figure 5. FTIR absorbance spectra in wavenumber range 600-3000 cm$^{-1}$, for PEO-PVP+NaPF$_6$ (Ö/Na$^+$), (a) 0, (b) 2, (c) 4, (d) 6, (e) 8 and (f) 10.

In the present system two interactions are possible for the cation, (i) electron rich ether group of PEO, (ii) C=O of PVP. When salt is added to the polymer blend then it will approach the suitable coordinating site and will alter the environment of the polymer chain. From the Figure 5 it is clearly visible that the C=O shows minor shift and decrease of intensity in peak position. The C-O-C stretching vibration band of PEO shows peak splitting in addition to salt in symmetric an asymmetric C-O-C stretching. Also, the shift in the wavenumber is noticeable which provides us strong evidence that the cation is going to coordinate with the electron-rich ether group of PEO.

*Polymer-ion interaction*

The addition of salt in the polymer blend alters the peak position, intensity and peak shape which demonstrates that the polymer salt complex formation occurs (Figure 6). This provides evidence of disruption of the polymer chains arrangement and transition from crystalline to amorphous region. Also, with the increase of the salt loading in the blend PEO-PVP matrix, the C-O-C stretching (amorphous) mode located at 1100 cm$^{-1}$ and C-H stretching mode 2850-2950 cm$^{-1}$ depicts suppression and broadening in the mode. This suppression and broadening of bands reveal the structural modification which alters the ordered arrangement and leads to reconstructions of molecular structures. It can be concluded from the above investigation that the observed shift in the positions, change in peak intensity of vibration bands suggest that the complex formation occurs on addition of salt. This is further confirmed by the impedance and the transport studies.

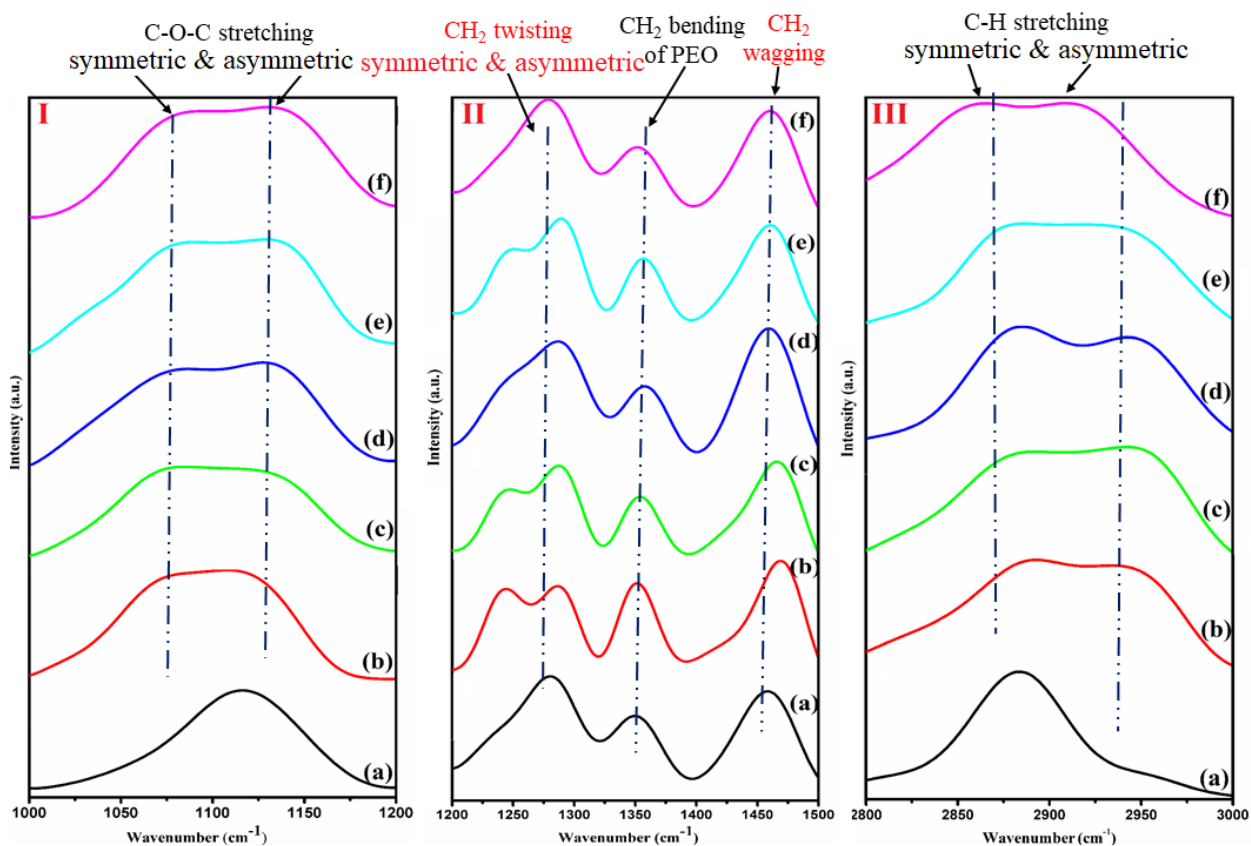

Figure 6. FTIR spectra for PEO-PVP+NaPF$_6$ (Ö/Na$^+$)= (a) 0, (b) 2, (c) 4, (d) 6, (e) 8, (f) 10 **(I)** C-O-C stretching mode, **(II)** CH$_2$ twisting/wagging/bending mode, **(III)** C-H stretching mode.

*ion-ion interaction*

The important investigation that leads to better understanding of complex formation is ion-ion interaction which will be examined by anion vibration mode (PF$_6^-$). The reason behind the investigation centered toward anion mode is that cation ($Na^+$) is IR inactive. Since the envelope of $PF_6$ anion seems asymmetric, deconvolution of the corresponding peak is done using Voigt Area function (in Peak Fit Software) to examine the free anion ($PF_6^-$) and ion pair ($Na^+$------$PF_6^-$) contribution in wavenumber region ~800-900 $cm^{-1}$. (Figure 7 a-e). Also, the baseline correction was done prior to deconvolution. It can be noticed from Figure 7 that addition of salt in the polymer blend leads to change in peak intensity and asymmetry starts appearing. The symmetry of anion is reduced from $O_h \rightarrow C_{3v}$ after interaction of cation [65-66]. The deconvolution pattern gives two type of vibration modes due to asymmetry, one at lower wavenumber side is attributed to 'free anion' vibration ($PF_6^-$), while at higher wavenumber side to the 'ion pairs' mode ($Na^+$---$PF_6^-$) [64]. A quantitative estimation of the fraction of free anion and ion pair was estimated from the area of deconvoluted peaks assigned to particular ions using following equation 3.

$$\begin{cases} \text{The fraction of the free anion (\%)} = \dfrac{A_{free}}{A_{free} + A_{pair}} \\ \qquad\qquad\qquad and \\ \text{Fraction of ion pair (\%)} = \dfrac{A_{pair}}{A_{free} + A_{pair}} \end{cases} \quad (3)$$

Here, $A_{free}$ is the area representing free ion peak and $A_{pair}$ area of the peak representing ion pair peak.

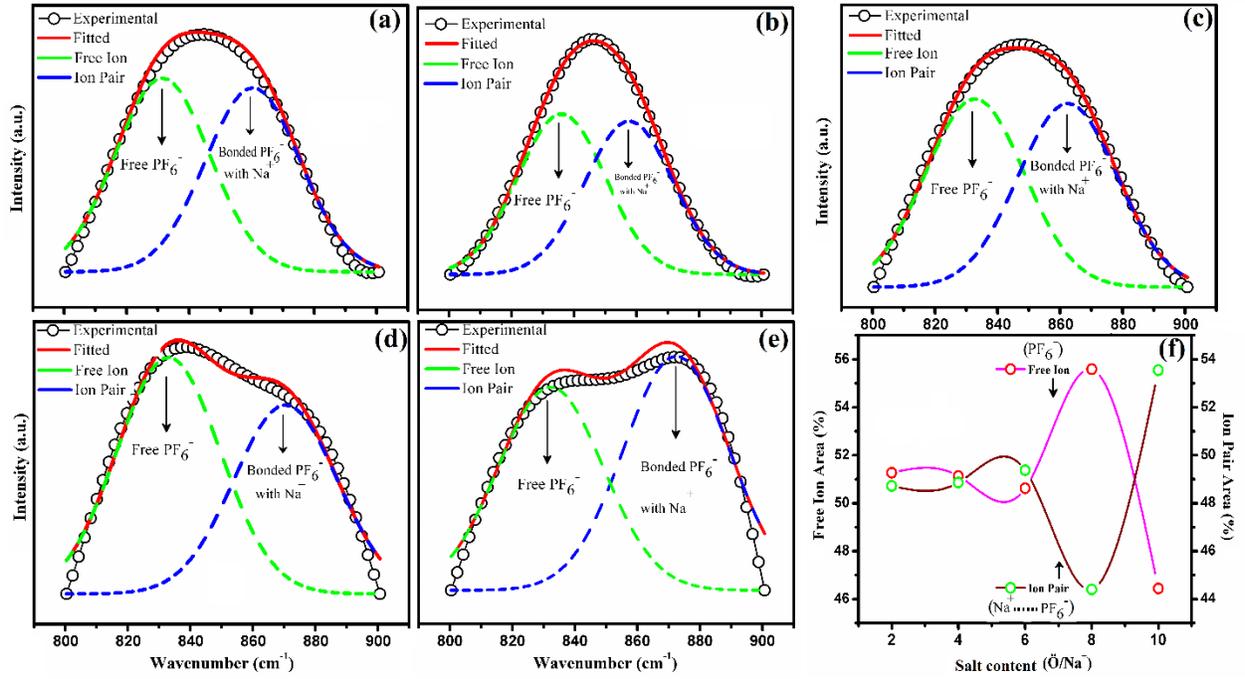

**Figure 7**. Deconvolution of the $PF_6^-$ vibration mode in the wavenumber range 800-900 $cm^{-1}$ for PEO-PVP+$NaPF_6$ (Ö/$Na^+$), (a) 2, (b) 4, (c) 6, (d) 8, (e) 10 using Voigt area function. (f) Salt concentration dependence of free anion and ion pair area.

A relative comparison of estimated corresponding free ion area and ion pair area is summarized in Table 3. As ion transport in polymer electrolyte is linked with polymer flexibility that is influenced by cation coordination and a number of free charge carriers, a fraction of free ions and ion pairs affect the ion transport. It is observed from the Table 3 that free ion contribution increases with salt concentration and the highest number of free ion carriers are generated for an optimum concentration which evidences the highest ionic conductivity, as shown later. This confirms that the complete salt dissociation achieved for an optimum concentration (Figure 7f) and corresponding to which ion pair exhibits a minimum. Further, a decrease in free ion area is responsible for the decrease of ionic conductivity which will be elaborated in the following section.

Table 3. The peak position of deconvoluted free ion and ion pair peak of SPE films.

| Sample code | Free ion | | Ion pair | | Corr. coff. ($r^2$) |
|---|---|---|---|---|---|
| | Area (%) | Wavenumber ($cm^{-1}$) | Area (%) | Wavenumber ($cm^{-1}$) | |
| SPE 2 | 51.27 | 832 | 48.72 | 860 | 0.992 |
| SPE 3 | 51.13 | 836 | 48.86 | 857 | 0.998 |
| SPE 4 | 50.62 | 833 | 49.37 | 862 | 0.994 |
| SPE 5 | 55.59 | 833 | 44.40 | 869 | 0.979 |
| SPE 6 | 46.44 | 833 | 53.55 | 870 | 0.954 |

**Electrochemical analysis**

*Impedance analysis*

The characteristic parameter of a solid polymer electrolyte is the ionic conductivity and it must be analogous to the existing technology consisting of the plasticized liquid polymer electrolyte. The impedance analysis of the SPE films has been carried out by sandwiching them between two stainless steel discs which act as blocking electrodes for cation/anion under an applied electric field. The cell assembly is shown in the inset of Figure 8a. The impedance plot for various salt concentration is shown in Figure 8a. The log-log presentation of impedance plots is chosen over the traditional plot for better clarity and comparison of the different impedance plots using single plot in a unique way. The enthusiasm behind this representation was from the superiority of log plots as elaborated by Jonscher [68-69].

In the high-frequency region of the complex impedance plots, there is a semi-circular arc for polymer blend which indicates that the conduction is mainly due to ions. In the low-frequency region, the presence of the inclined line suggests the effect of the blocking electrodes. This is evidence of the presence of capacitive nature and absence of electronic conductivity [70-71]. The dip in the plot associated with the minima in $Z''$ provides us the value of bulk resistance on the real axis. The nature of all curves is almost identical. Addition of salt in the polymer blend shifts the dip in the curve towards lower impedance side which can be correlated with the enhancement of the ionic conductivity calculated by equation 1. The highest ionic conductivity was $5.92 \times 10^{-6}$ S $cm^{-1}$ obtained for sample SPE 5 which is $Ö/Na^+ = 8$ in the PEO-PVP blend. A two-fold conductivity enhancement was achieved at room temperature as compared to the blend polymer electrolyte. This composition of blend polymer with salt ($Ö/Na^+ = 8$) has been referred to as the optimum conducting composition (OCC). As the conductivity is linked to a number of charge carriers and the mobility, the addition of salt alters the polymer chain arrangement which enhances the amorphous content and cation migration becomes easier. The observed room temperature ionic conductivity in the present work is higher as compared to previously reported values [52, 56, 72]. It is also observed that the conductivity increases with the addition of salt and decreases with the further addition of salt. This decrease in conductivity may be associated with the incomplete dissociation of salt. This reduces the number of charge carriers and ion pair formation as evidenced by the deconvolution of FTIR. Also, sample SPE 2 displays another maximum in electrical conductivity. Figure 8b displays the fitted impedance plot of OCC (Sample 5) with an equivalent circuit using the $Z_{SimpWin}$ software. The equivalent circuit shown in inset consists of the constant phase element ($Q_1$) in parallel with resistance and in series with another constant phase element ($Q_2$).

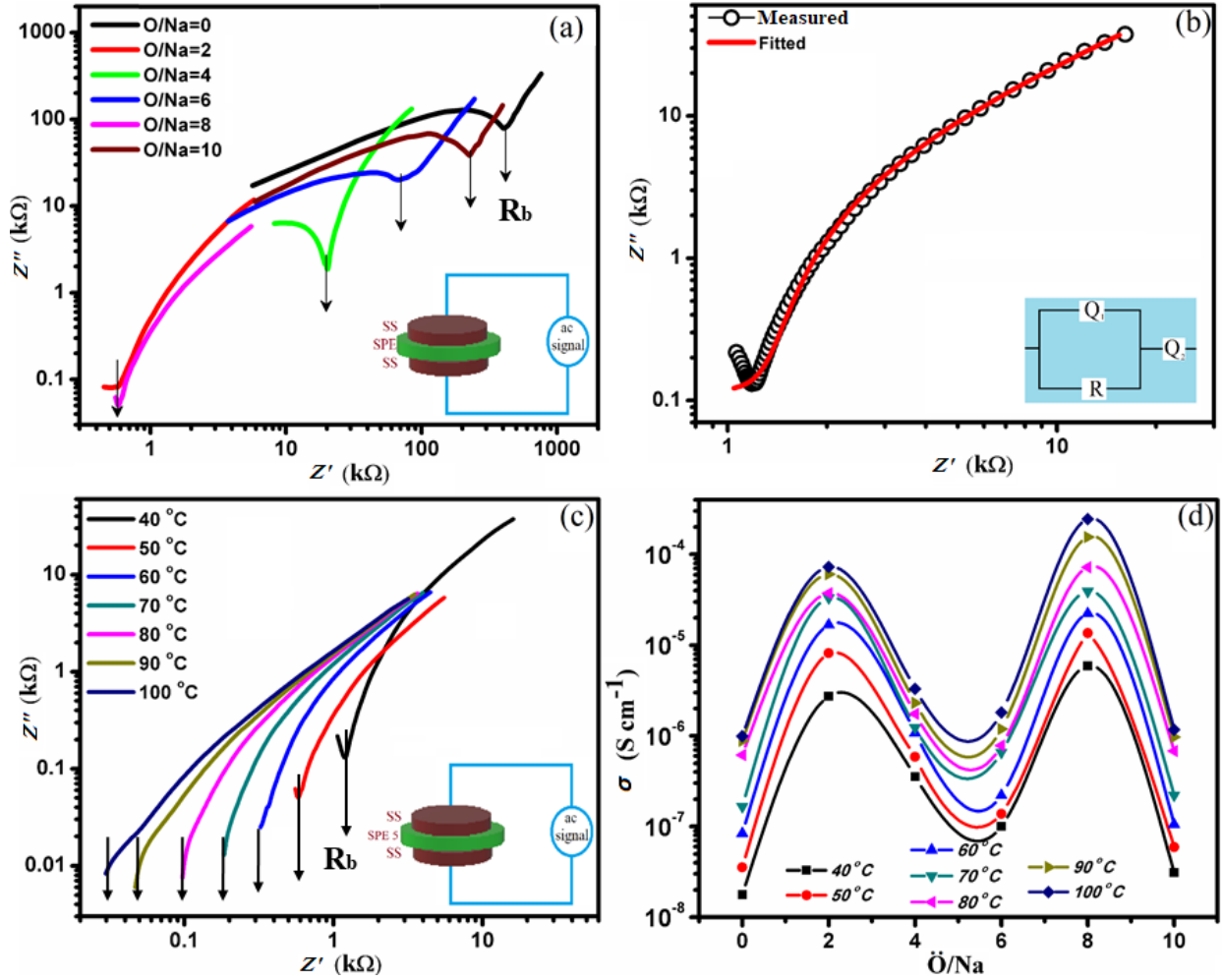

Figure 8. **(a)** Room temperature log-log complex impedance plots for different salt concentration (Ö/Na$^+$ = 2, 4, 6, 8, 10). Inset of (a) depicts cell configuration, **(b)** fitting of impedance plot for SPE 5. Inset of (b) depicts fitted circuit element, **(c)** log-log impedance plot of SPE 5 at different temperature (40 °C – 100 °C) and **(d)** variation of ionic conductivity with salt concentration (Ö/Na$^+$ = 2, 4, 6, 8, 10) at different temperature (40 °C–100 °C).

Table 4. Conductivity variation with temperature for different salt content.

| Sample code<br>Temperature | SPE 1 | SPE 2 | SPE 3 | SPE 4 | SPE 5 | SPE 6 |
| --- | --- | --- | --- | --- | --- | --- |
| 40 °C | 1.77×10$^{-8}$ | 2.74×10$^{-6}$ | 3.55×10$^{-7}$ | 1.01×10$^{-7}$ | 5.92×10$^{-6}$ | 3.11×10$^{-8}$ |
| 50 °C | 3.54×10$^{-8}$ | 8.19×10$^{-6}$ | 5.87×10$^{-7}$ | 1.38×10$^{-7}$ | 1.36×10$^{-5}$ | 5.94×10$^{-8}$ |
| 60 °C | 8.40×10$^{-8}$ | 1.67×10$^{-5}$ | 1.08×10$^{-6}$ | 2.21×10$^{-7}$ | 2.24×10$^{-5}$ | 1.05×10$^{-7}$ |
| 70 °C | 1.66×10$^{-7}$ | 3.32×10$^{-5}$ | 1.23×10$^{-6}$ | 6.56×10$^{-7}$ | 3.92×10$^{-5}$ | 2.22×10$^{-7}$ |
| **80 °C** | 6.15×10⁻⁷ | 3.74×10⁻⁵ | 1.74×10⁻⁶ | 6.43×10⁻⁷ | 7.22×10⁻⁵ | 6.79×10⁻⁷ |
| **90 °C** | 8.82×10⁻⁷ | 6.05×10⁻⁵ | 2.31×10⁻⁶ | 1.19×10⁻⁶ | 1.55×10⁻⁴ | 9.69×10⁻⁷ |
| **100 °C** | 9.89×10⁻⁷ | 7.29×10⁻⁵ | 3.29×10⁻⁶ | 1.81×10⁻⁶ | 2.46×10⁻⁴ | 1.17×10⁻⁶ |

**Temperature dependence of ionic conductivity**

The ionic conductivity of the prepared SPE films was measured in the temperature range of 40 °C to 100 °C. The ionic conductivity was highest for sample SPE 5 with composition PEO-PVP+Ö/Na$^+$ = 8. As temperature rises, it enhances the polymer chain flexibility which leads to faster ion mobility, and hence ionic conductivity. So, further insight toward understanding the effect of temperature, impedance was measured at a different temperature. The measured impedance plots of the SPE 5 sample at different temperatures are presented in Figure 8c. It can be noticed from Figure 8c that the dip in the curve shift towards the lower impedance side which indicates the lowering of bulk resistance and hence the enhanced ionic conductivity. This enhancement in the ionic conductivity with an increase of temperature may be associated with the increased polymer chain flexibility and a number of free charge carriers. Also, the activation of free charge carriers between the coordinating sites and the segmental motion of the polymer chains may occur due to increased free volume [72-73]. This increased flexibility of polymer chains indicates the increase in the ion mobility which promotes the smoother ion migration.

Figure 8d depicts the variation of ionic conductivity with salt content at a different temperature. It can be concluded from Figure 8d that the ionic conductivity increases with the temperature and is maximum at the highest temperature. The highest ionic conductivity achieved was 5.92×10⁻⁶ S cm⁻¹ at 40 °C and increased to 2.46×10⁻⁴ S cm⁻¹ at 100 °C (Table 4). A comparison of ionic conductivity for a various blend polymer matrix with different salts is represented in Table 5 and it can be concluded that the present blend polymer matrix has improved conductivity more than the other reports known from literature till date. It validates the use of synthesized BPE for application in sodium-based energy storage devices. Figure 8d displays the temperature dependent ionic conductivity of the investigated polymer electrolyte in the temperature range of 40 °C to 100 °C. Here, the ionic conductivity seems to be increased with the temperature owing to the enhancement of the polymer flexibility which increases the cation (Na$^+$) mobility. Almost a linear variation in conductivity is observed with temperature. For the lower salt concentration (Ö/Na$^+$ = 10) only a minor increase in conductivity is observed and may be due to the availability of less number of free charge carriers. Further, the conductivity increases up to Ö/Na$^+$ = 8 and after that is followed by a decrease for the Ö/Na$^+$ = 4, 6. This may be due to the optimum value of Na w.r.t. number (8) of the polymer chain and continuous hanging of anion in the polymer chain. While for the very high salt content again conductivity is increased that may be due to the effective role played by anion along with cation in charge transport Further increasing the salt concentration lowers trend of electrical conductivity due to strong ion pairing of Na/PF$_6$.

Table 5. Comparison of the ionic conductivity of various sodium and lithium salts.

| Polymers | Salt | Ionic conductivity | Temperature | Reference |
|---|---|---|---|---|
| PEO-PEMA | NaClO$_4$ | 6.77×10⁻⁷ S cm⁻¹ | 30 °C | [74] |
| PEO-PVP | NaIO$_4$ | 1.56×10⁻⁷ S cm⁻¹ | RT | [75] |
| PEO-PVP | NaF | 1.19×10⁻⁷ S cm⁻¹ | RT | [46] |

| PVC-PEMA | NaIO$_4$ | $10^{-8}$ S cm$^{-1}$ | RT | [76] |
|----------|----------|----------------------|------|---------|
| PEO-PVP  | LiClO$_4$ | $3.77\times10^{-6}$ S cm$^{-1}$ | 40 °C | [56] |
| PEO-PVP  | NaBr     | $1.90\times10^{-6}$ S cm$^{-1}$ | RT | [54] |
| PEO-PVP  | NaPF$_6$ | $5.92\times10^{-6}$ S cm$^{-1}$ | 40 °C | Our work |
| PEO-PVP  | NaPF$_6$ | $2.46\times10^{-4}$ S cm$^{-1}$ | 100 °C | Our work |

The enhancement of the ionic conductivity is attributed to the salt dissociation owing to the polymer-ion and ion-ion interaction. The salt dissociation results in the release of a free number of ions and is highest for the content which shows conductivity maxima as evidenced by the FTIR deconvolution. The first maxima is associated with the cation migration as an anion is hanged to the polymer backbone. So, the segmental motion of the polymer chain plays an effective role in enhancing the ion transport. Then another maximum is achieved at high salt content which is associated with the anion migration. As at high salt content both cation and anion increase, former one gets trapped via the ion crosslinking formation, while anion stays out the cation path due to high concentration. This results in the disorder in the polymer matrix and anion starts migrating via the polymer backbone having CH$_2$ group. Since anion bearing larger size and mass so it demonstrates comparative conductivity with the cation. Further, explanation of the conductivity maxima is explained by the transport parameters measurement and the mechanism termed as two percolation peak mechanism in the upcoming section.

**Conductivity stability and ion transference number**

To highlight the suitability of solid polymer electrolyte at high temperature (100 °C), conductivity monitoring is done for the optimized PEO-PVP+NaPF$_6$ (Ö/Na$^+$ =8) highly conducting system. The conductivity can still be maintained stably up to 160 h (approx. 7 days) as shown in Figure 9a. The above advantages of the prepared solid polymer electrolyte can serve as a suitable candidate for long-term safe solid-state sodium ion battery.

The fraction of current carried by ions in the prepared SPE is important to analyze. Now, the ion transference number measurement was performed using Wagner's DC polarization technique. As SPE 5 system has optimum conductivity for practical application in energy storage devices, we performed the ion transference number measurement for the PEO-PVP+NaPF$_6$ (Ö/Na$^+$ =8) system with an applied dc bias ~50 mV across the configuration SS|SPE 5|SS (Figure 9b). The plot shows initially high current ($i_t$) and drops sharply with an increase of time followed by a steady state associated with current $i_e$. The former one is a contribution from both ions and electrons while the later one is an only electronic current contribution. For the perfect ionic conductor, the value of ion transference number is supposed to be unity. We have measured the ion transference number ($t_{ion}$) for the highest conducting sample and is found to be 0.997 using equation 2.

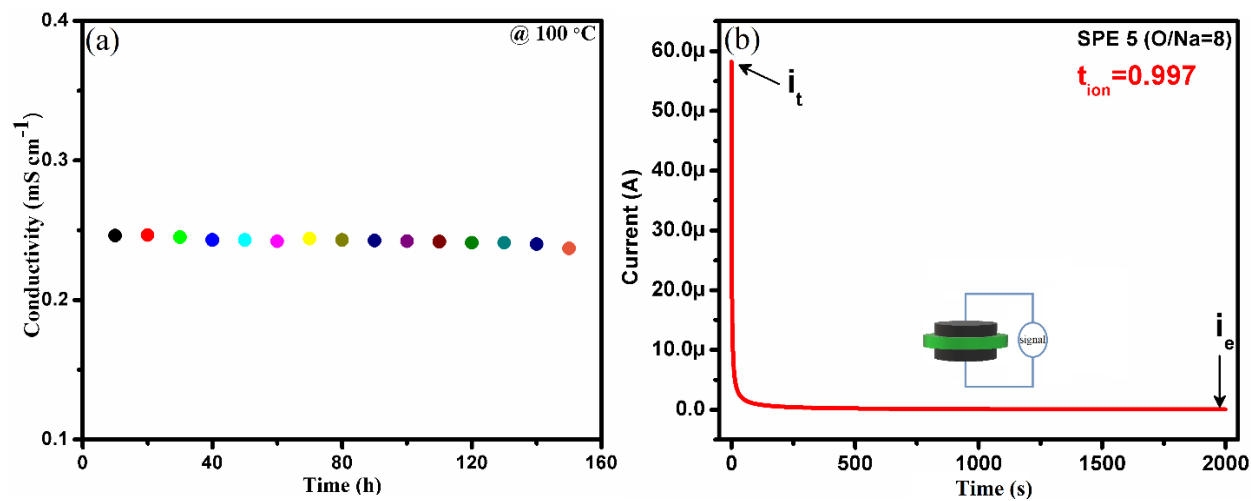

**Figure 9.** (a) Conductivity stability of the PEO-PVP+NaPF$_6$ (Ö/Na$^+$ =8) at a high temperature of 100 °C and (b) variation of current with respect to time across the cell; SS|SPE 5|SS.

On the application of electric field on the cell configuration SS|SPE|SS, both ions and electrons respond to the applied field. The stainless steel (SS) electrodes prevent the ion flow to an external circuit by creating a zone of mobile ions on the electrode/electrolyte interface. Then further ion migration is constrained by this zone, after some time it dominates over the applied field and concentration polarization is achieved. Now, the current starts decaying due to the balance of drifted ions and diffused ions. As a result of this, concentration polarization zone formed at electrode/electrolyte interface increases the interfacial resistance and ionic current is blocked now while permits electronic current. So, after some time only electronic current is dominating. This value is close to value of perfect ionic conductor and it concludes that the present system is ion dominating with most of the charge transfer (~99 %) by ions with negligible electron contribution [77-78].

**Thermal activation energy measurement**

Another important parameter to investigate the ion migration is activation energy ($E_a$). It depicts the energy associated with the defect formation and the ion migration simultaneously. The value of activation energy basically tells about the favorable environment and the conditions for the ion migration for smoother ion transport. The high value of activation energy may be due to the requirement of high energy for ion migration [79-80]. Now, all the plots were fitted with the Arrhenius equation to obtain the value of activation energy ($E_a$) from the slope of linear portion (Figure 10 a). From the plot, it can be concluded that all SPE shows perfect agreement with the Arrhenius equation (the solid red line is the fitted equation). From the results, it was obtained that the SPE system with the highest conductivity (SPE 5) sample has conducting pathways of the lowest activation energy as compared to other blend polymer salt complexed electrolyte (Figure 10b). So, ionic conductivity depends on the activation energy of the free ions directly and smaller the activation energy, the smoother the cation migration. This lowering in the activation energy is owing to the increase of flexibility of polymer chains for this concentration which makes a favorable conductive path for the ions. All SPE shows the value of activation energy in the range of 0.2 eV to 0.3 eV and is in desirable range for the fast solid state ionic conductor. Another point to be noted here is that the activation energy values are in absolute agreement with the free ion area obtained from FTIR and the impedance study.

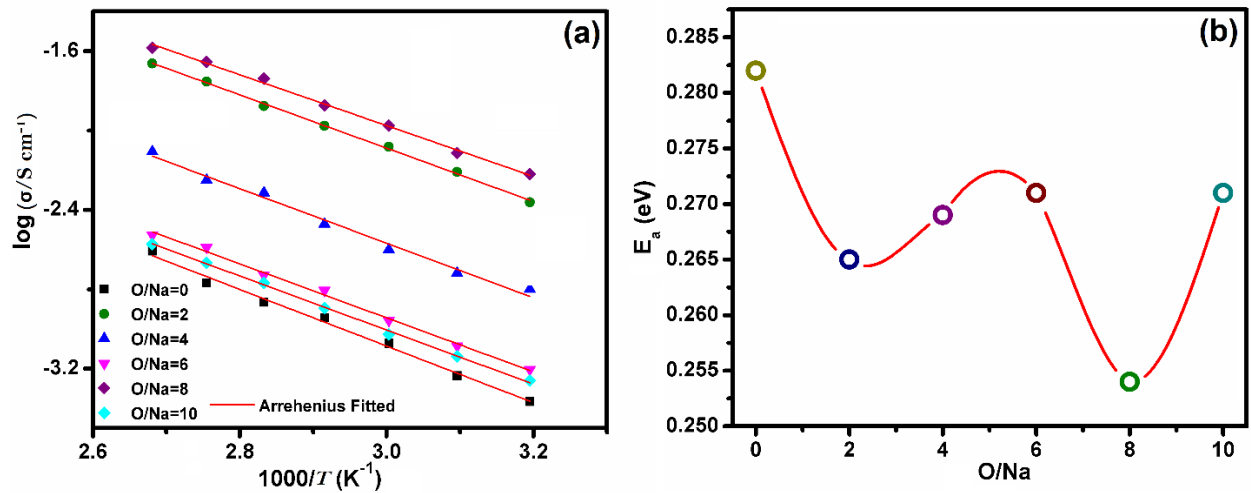

**Figure 10.** (**a**) Arrhenius plot and (**b**) activation energy plots for PEO-PVP blend polymer electrolyte with different salt concentration ($\mathrm{\ddot{O}/Na^+} = 2, 4, 6, 8, 10$).

**Transport parameters**

The desirable property of a solid polymer electrolyte is high ionic conductivity ($\sigma$) which is directly linked with number density ($n$), mobility ($\mu$) of charge carriers, viscosity ($\eta$) and diffusion coefficient ($D$) for any plastic separator stands for electrolyte cum separator. As the electrical conductivity value of solid polymer electrolyte films is directly dependent on a number of free charge carriers and mobility ($\sigma = nq\mu$), the FTIR spectroscopy was used to evaluate the parameters using the equation reported somewhere [81]. FTIR deconvolution was done to determine the percentage area of free ion and ion pair and the areas are plotted as a function of salt content (Figure 7d). Figure 11 displays the variation of mobility ($\mu$), viscosity ($\eta$) and diffusion coefficient ($D$) of charge carriers against the different salt content. It reveals the one-to-one correspondence between charge carrier mobility ($\mu$) and diffusion coefficient ($D$) along. Both the mobility and diffusion coefficient of charge carriers increases with addition of salt. The maximum for both was achieved for the optimum conductivity value (PEO-PVP+NaPF$_6$ ($\mathrm{\ddot{O}/Na^+}$ =8)). Also, viscosity of polymer chains influence the charge migration so it was investigated and follows inverse trend as of mobility or conductivity. The maxima in the conductivity or mobility is associated with minima in the viscosity. This concludes that the viscosity of polymer chain reduces with addition of salt and leads to faster ion migration [82]. This result is also in good agreement with the associated FTIR ion-ion interaction and conductivity studies.

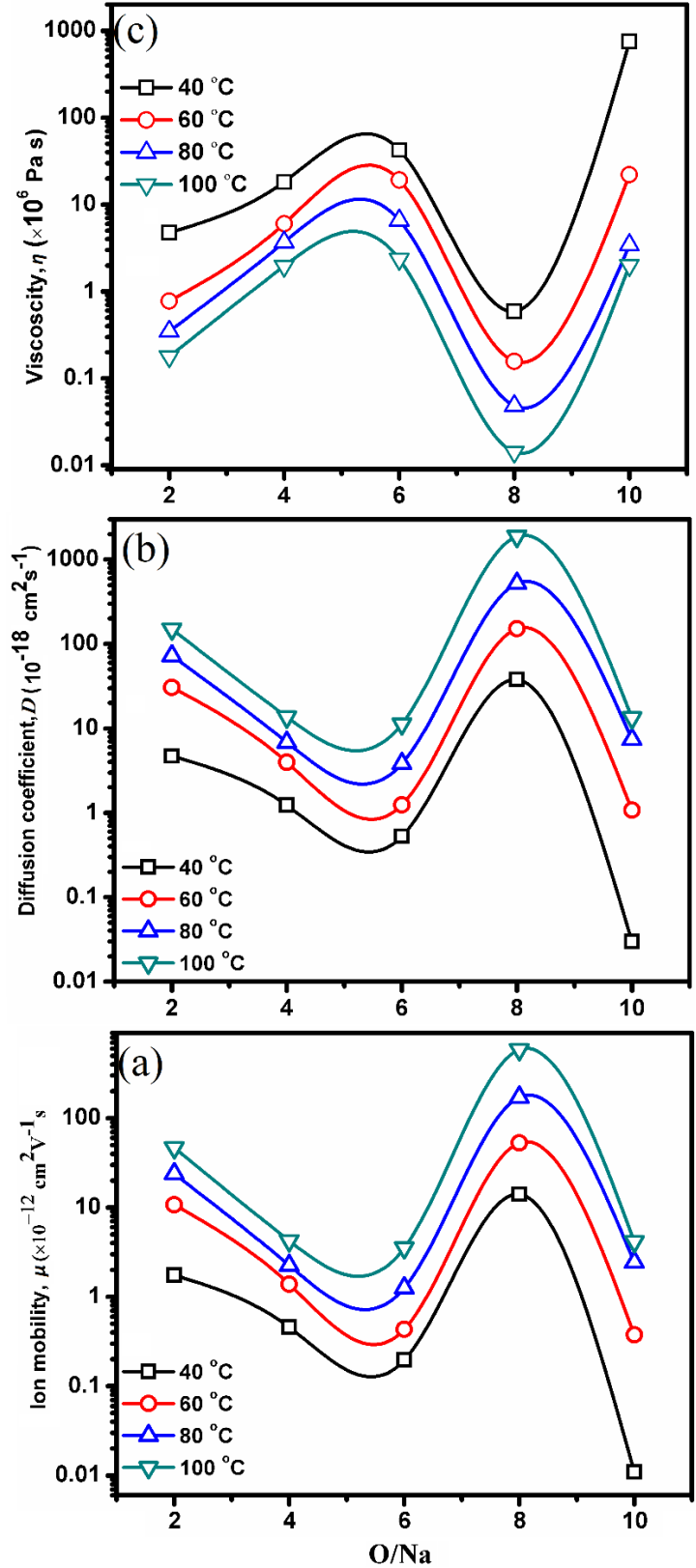

**Figure 11.** The plot in a variation of (a) mobility ($\mu$) of charge carriers, (b) viscosity ($\eta$) and (c) diffusion coefficient (*D*) against the salt content.

**Differential scanning calorimetric (DSC) analysis**

In case of polymer electrolytes, various phase transitions occur with temperature and provide crucial information to explain the enhancement of the electrical properties. So, DSC was performed (Figure 12) to measure the glass transition temperature ($T_g$), melting temperature ($T_m$), crystallinity ($X_c$) and their variation with salt content is shown in Table 6. It needs to be mention here that the glass transition is linked to the polymer chain flexibility and ion mobility. The low value of $T_g$ indicates faster ionic transport due to the increased polymer chain segmental motion. The blend polymer electrolyte shows $T_g$ about -69.02 °C and the addition of salt in the polymer blend evidence the shift toward lower temperature. Further addition of salt in the blend polymer electrolyte demonstrates little change in the $T_g$ with the lowest value for the SPE 5 (O/Na=8) followed by the SPE 2 (O/Na=2). The shift on the $T_g$ is attributed to the reduction of the covalent bonding between the polymer chains due to ion penetration. The almost same trend is followed by the peak associated with melting temperature ($T_m$) where polymer gets melted. The shift of peak toward lower temperature indicates the increased ionic conductivity and both $T_g$ & $T_m$ supports this approach [83-84].

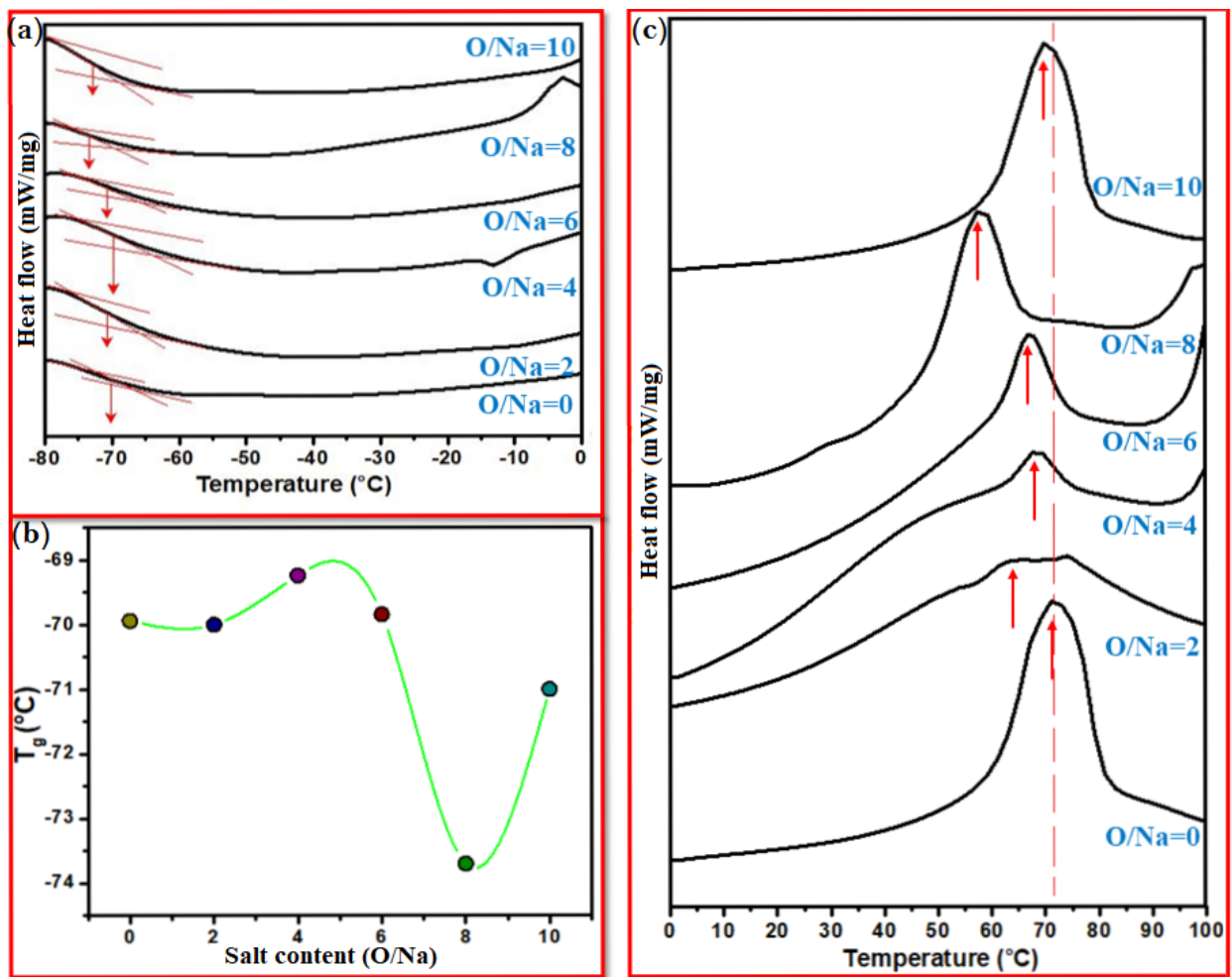

**Figure 12.** DSC thermograms for the solid polymer electrolytes, (a) DSC trace, (b) variation of glass transition temperature against salt content and (c) variation of the melting peak. Approximate O/Na ratio is also given for each sample.

The crystallinity was calculated using equation 3 and it shows the crystallinity minima for the SPE 2 system. Although the crystallinity is lowest for the blend polymer electrolyte with O/Na=2, while conductivity was highest for the O/Na=8 based polymer electrolyte. The crystallinity with the addition of optimum salt content is much smaller in comparison to the blend polymer electrolyte (without salt). This indicates that the addition of salt disrupts the crystalline arrangement of the polymer chain by eliminating the covalent bonding the polymer chains ($Na^+$---O- and $PF_6^-$----$H_2C$-). So, the enhanced amorphous content is achieved, which favors the faster ion migration.

Table 6. Thermal properties of the blend polymer with different salt content.

| Sample code | $T_g$ (°C) | $T_m$ (°C) | $X_c$ (%) |
|---|---|---|---|
| O/Na=0 | -69.02 | 71.71 | 49.36 |
| O/Na=2 | -72.03 | 68.21 | 1.29 |
| O/Na=4 | -70.23 | 67.25 | 2.26 |
| O/Na=6 | -70.42 | 66.89 | 4.72 |
| O/Na=8 | -77.21 | 59.31 | 5.21 |
| O/Na=10 | -72.89 | 70.78 | 47.56 |

The enhanced amorphous content is also linked to the increased free volume that provides the smoother path to the cation. It can be concluded that the lowering of the $T_g$ and $T_m$ value along with crystallinity reduction with the addition of salt provides sufficient evidence for the enhancement in the conductivity.

**Atomic force microscope (AFM) analysis**

AFM has been used to study the morphological changes after addition of salt in blend polymer electrolyte (BPE). The two-dimensional image of the polymer blend (without salt) and polymer blend with $Ö/Na^+ = 8$ having optimum ionic conductivity are shown in Figure 13. From Figure13 a & b, it is observed that the average roughness value of the polymer blend is about 80 nm. The surface roughness factor plays important role in the enhancement of ionic conductivity and linked with amorphous content. When salt is added to the polymer blend then the severe surface modification indicates that the salt plays an active role as shown in Figure 13 c & d. Addition of salt reduces the crystallinity of the blend polymer electrolyte and makes surface smoother which is favorable for faster ion conduction [85]. Now, the surface roughness value increases to about 130 nm. This increase in roughness value indicates the complex formation of a polymer blend with salt and amorphous content improved [86]. This increase in the roughness parameter and amorphous content is responsible for the enhancement of ionic conductivity and promotes ion migration. It can be concluded that the addition of salt modifies the structure of the blend polymer electrolyte (BPE) and is in close agreement with the XRD, FESEM and Impedance study.

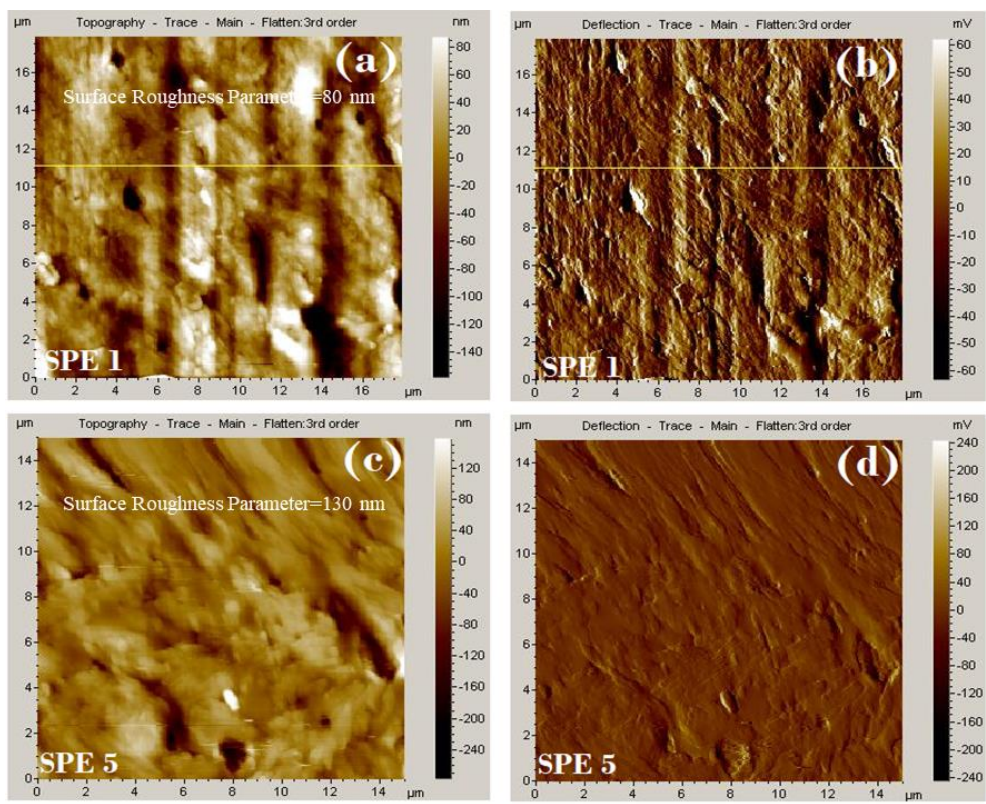

**Figure 13.** Two dimensional **(a & b)** topography and deflection image of PEO-PVP and **(c & d)** Topography and deflection image of PEO-PVP+NaPF$_6$ (Ö/Na$^+$ = 8).

**Thermo-gravimetric analysis (TGA)**

Thermal stability of a battery device is important to avoid the material decomposition and explosion during cell operation. So, the thermal stability was investigated by TGA to check the safety window of solid polymer electrolyte [87]. The thermo-grams of PEO-PVP blend and PEO-PVC+NaPF$_6$ with Ö/Na$^+$ =2 & 8 are plotted in Figure 14 and thermogram is divided into three regions for a better understanding of various decomposition stages.

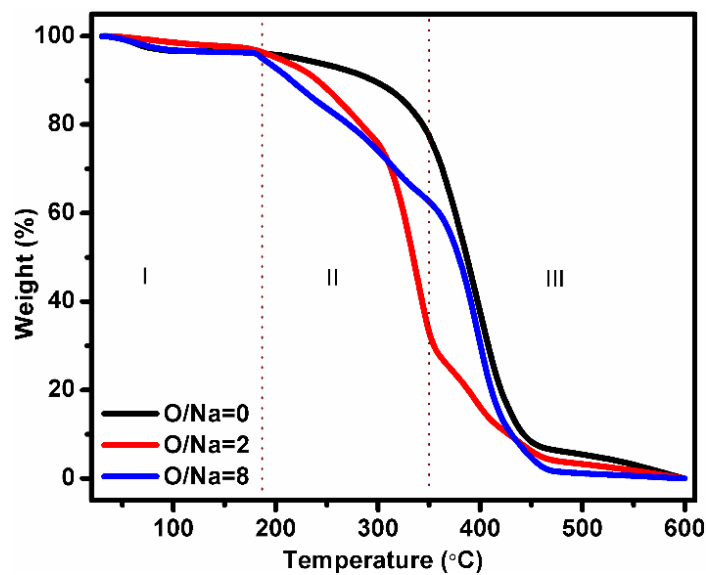

**Figure 14.** TGA curves of PEO-PVP blend, PEO − PVP + Ö/Na$^+$ = 2 and PEO-PVP+Ö/Na$^+$ =8.

In the *region 1* initially a small weight loss is observed at 60 °C -80 °C and that may be due to the evaporation of the solvent and the moisture content during loading of the sample [62, 88]. PEO-PVP blend with both salt stoichiometric ratio shows identical thermal stability.

Now, in *region 2* PEO-PVP blend shows a small weight loss of approx. 15 %. In PEO-PVP blend without salt one step loss is observed while after addition of salt one additional step is clearly visible. This multi-step process confirms that the present system is BPE [89]. Addition of salt alters the thermal stability and for higher salt content (Ö/Na$^+$ =2) more weight loss is observed as compared to Ö/Na$^+$ =8. Here, salt is acting as an opponent and reduce thermal stability of overall solid polymer electrolyte. This may be due to the increased flexibility of polymer chains. Now, less energy is required to disrupt the bonding and reduction of thermal resistance leads to decomposition at a lower temperature. However, the addition of salt shifts the graph towards lower temperature only up to 70 °C -80 °C. The rapid weight loss in various steps confirms the degradation of the sample beyond 300 °C. In the *region 3*, the polymer matrix constituents such as a polymer, salt start to degrade and weight loss are maximum. The solid polymer electrolyte cannot be completely decomposed even when the temperature reaches 600 °C. In case of polymers, at high temperature two types of decomposition process are known, (i) chain de-polymerization or unzipping, (ii) random decomposition. The former one is just the release of the chains from weak link while later one occurs via chain rupturing at random points. Both processes contribute to loss of mass [54]. The plateau region or almost zero loss region confirms the thermally stable up to approx. 200 °C and is enough to fulfill the demand of electrolyte in energy storage/conversion devices. Obviously, we can claim from these observations that these SPE can be operated up to 200 °C and are preferred in the SIB as its operating temperature is normally below 100 °C.

**Electrochemical stability window**

Electrolyte stability window (ESW) enables us to obtain the maximum operating neutral voltage range with an applied potential of the polymer electrolyte for device applications. Linear sweep voltammetry is performed to obtain the ESW in the configuration of SS/SPE5/SS at 40 °C and 80 °C within the voltage range of 0-5 V at a scan rate of 10 mV/s. From Figure 15 it can be noticed that initially the current remains in steady state and then shows a sharp increase that is attributed to the electrolyte decomposition at inert electrode interface [90-92].

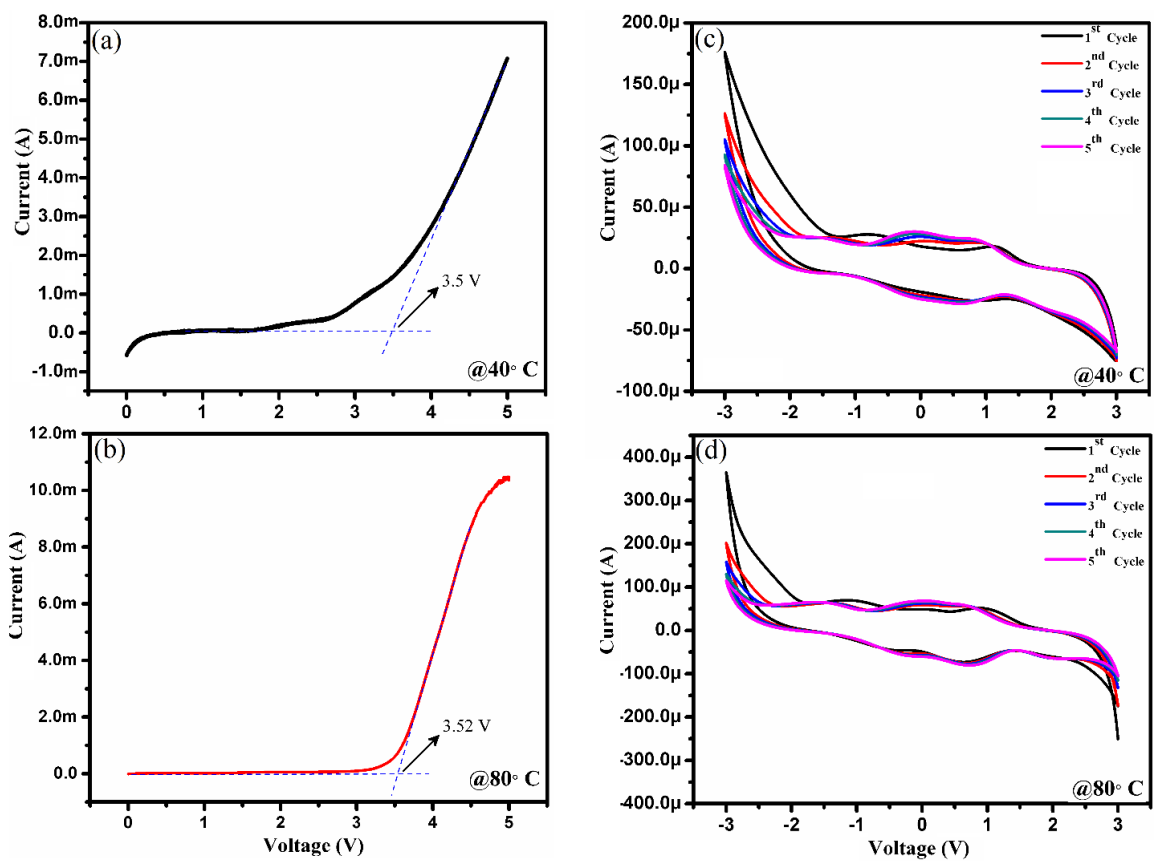

**Figure 15.** Linear sweep voltammetry of SS/SPE 5/SS, (a) 40 °C, (b) 80 °C and Cyclic voltammetry curves of the system SS/SPE 5/SS, (c) 40 °C and (d) 80 °C.

The voltage window is up to at least 3.5 V for the SPE 5 based system (Figure 15a). Furthermore, the high-temperature stability (80 °C) of the SPE 5 system is investigated. From Figure 15b, it is clearly visible that the voltage window is stable up to 3.5 V which safeguards the application of prepared solid polymer electrolyte at high temperature.

The cyclic voltammetry (CV) study of the SPE 5 system was measured in the voltage range -3 to 3 V with a scan rate of 10 mV/s for 5 cycles (Figure 15c & d). The configuration for the measurement was SS/SPE 5/SS. From Figure 15c it can be concluded that the stability of polymer electrolyte remains even after 5 cycles and this evidences the applicability of polymer electrolyte for application in the voltage range of at least 5 V. Also, the repeatability in the curve confirms the voltage stability of the investigated system. Now, to test the high-temperature applicability of prepared polymer electrolyte, high-temperature voltage stability (80 °C) is measured by CV. Figure 15d shows both cyclic and thermal stability of the solid polymer electrolyte. Another noteworthy point is that absence of any cathodic/anodic peak for both low and high-temperature CV curve confirms the broad stability window for the long-term application. Overall, both LSV and CV support the use of prepared polymer electrolyte with long-term cycle stability and thermal stability for the solid state sodium ion batteries.

**Two peak percolation model/mechanism**

The enhancement of the ionic conductivity is further examined by proposing the model that indicates the individual role played by the polymer and the salt. This model is based on the results obtained by the FTIR, impedance study and the transport parameters. Figure 16 (a-g) depicts the ion transport mechanism. When the salt is added in blend polymer, various interactions occurs depending on the availability of sites for interaction. Figure 16a displays the four possible interactions as a general case, (i) F of anion with the H of PEO, (ii) P of anion may interact with N of PVP, (iii) F of anion with the H of PVP and (iv) P of anion may interact with O of PVP. Out of all four possibilities first and second seems to be feasible and more appropriate in the present case based on microscopic interaction among a different functional group of polymer and salt through FTIR result findings. The hydrogen bonding between the F of the anion with the H of PEO and anion interaction with the blend polymer is effective in the formation of the polymer salt complex. First of all, when the two polymers are dissolved then both PEO and PVP interact via the hydrogen bonding (Figure 16b&c). This formation of the blend was also confirmed by the DSC as single glass transition temperature was observed for the present system. Now when the salt is added to the polymer electrolyte then the salt get dissociated in the cation and anion. Now, cation gets coordinated with the ether group of the polymer chain as evidenced by the FTIR (Figure 16d). The anion gets coordinated with N of the PVP. This interaction alters the polymer chain arrangement and the viscosity of chain decreases. Now, when the addition of the salt is done in the blend polymer electrolyte then the migration of the ion occurs via the segmental motion of the polymer chains. The salt gets coordinated with the ether group of the host polymer while anion with the polymer backbone. The formation of new coordinating sites and elimination of previous sites leads to cation transport (Figure16e). The sufficient sites are to participate in the ion transport. With the increase of the salt content, dissociation of the salt increases as well as a number of free charges carriers. As conductivity is linked to the number of free charge carriers so conductivity is enhanced and first maxima are achieved (Figure 16e). Now, when the salt content is increased then there arises the possibility of the ion pair formation as observed by the FTIR deconvolution and decrease of free ion area is observed. Figure 16f shows the formation of ion pairs due to the absence of sufficient cording sites for cation coordination. This restricts the salt dissociation and conductivity decrease. Now, further increase of the salt content enhances the ion transport, the ion charges carriers are anions. As now the disorder in the polymer chain is more and number of anion are also large. As cation is of smaller size than the anion so it gets trapped in between the polymer chains via the ion-crosslinking that traps the cation and halts its migration. As it is well known that the anion is attached to the polymer backbone or resides outside the polymer chains, the anion starts participating in the conduction via the polymer backbone group and second maxima is achieved (Figure 16g). Further addition of the salt content in the blend polymer electrolyte leads to the ion-triplets formation and lowering of the conductivity is observed (Figure 16h). Actually, at high salt content, the segmental motion is not so effective and salt dissociation is not observed.

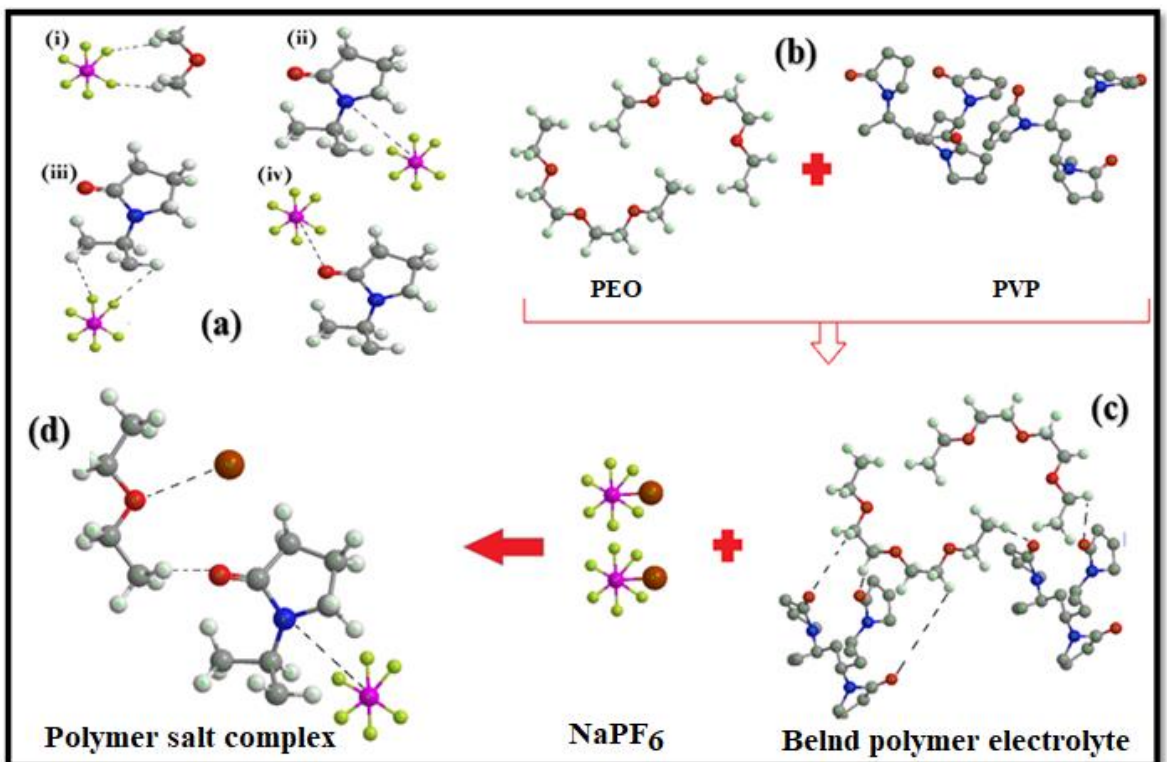

(a) (i), (ii), (iii), (iv)
(b) PEO + PVP
(c) Blend polymer electrolyte
(d) Polymer salt complex
NaPF$_6$

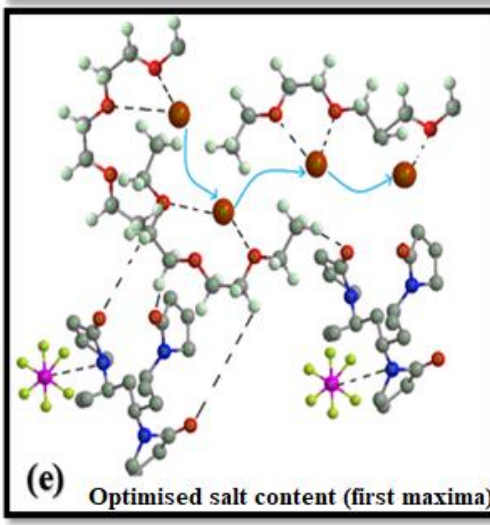

(e) Optimised salt content (first maxima)

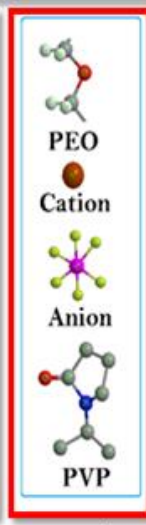

PEO
Cation
Anion
PVP

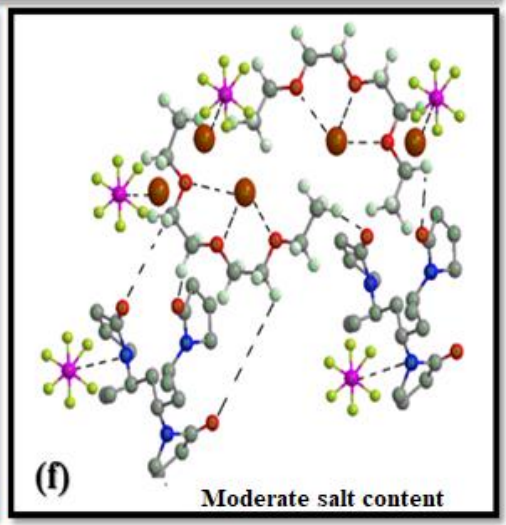

(f) Moderate salt content

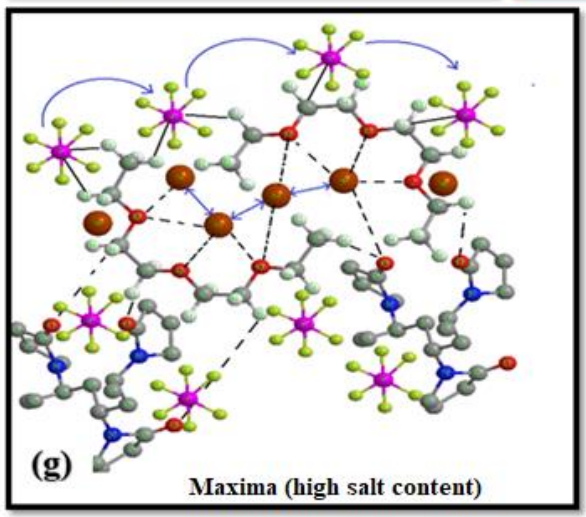

(g) Maxima (high salt content)

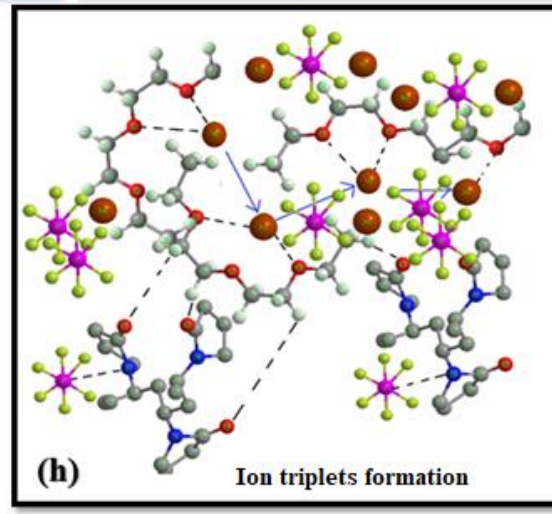

(h) Ion triplets formation

**Figure 16.** Proposed mechanism, (a) four possible interaction of the salt with polymer matrix, (b, c) blend polymer electrolyte formation, (c) polymer salt complex formation, and interaction at (d) optimum salt content, (e) moderate salt content, (f) high salt content and (g) very high salt content. The bold rectangle depicts the various active species in the interaction.

**Conclusions**

The flexible and freestanding blend solid polymer electrolyte films based on PEO-PVP complexed with $NaPF_6$ were synthesized by the standard solution cast technique. The reduction of peak intensity from the XRD evidenced the increase of the amorphous content and increased interchain spacing. FESEM micrograph displays the formation of the smoother surface with the addition of salt which indicates the enhanced amorphous content. AFM analysis also supports the enhancement of the amorphous content. FTIR confirmed the cation coordination with the ether group of the polymer chain and polymer salt complex formation explained in terms of the presence of polymer-ion, ion-ion interactions. The deconvolution of FTIR band associated with the anion reported that the free ion area is enhanced with the addition of salt which suggests the proper salt dissociation. The film PEO-PVP+$NaPF_6$ (Ö/$Na^+$ =8) exhibits highest ionic conductivity ~$5.92 \times 10^{-6}$ S cm$^{-1}$ at 40 °C and ~$2.46 \times 10^{-4}$ cm$^{-1}$ at 100 °C. The temperature dependent conductivity shows Arrhenius type behavior and a decrease in activation energy evidence the increased flexibility and hence the enhanced ionic conductivity. DSC also supports the enhancement of the ionic conductivity as both $T_g$ and $T_m$ shifts toward lower temperature. The high temperature (100 °C) conductivity monitoring is done for the optimized PEO-PVP+$NaPF_6$ (Ö/$Na^+$ =8) highly conductive system and the conductivity can still be maintained stably up to 160 h (approx. 7 days). The prepared polymer electrolyte film displays the smoother surface in addition of salt and a thermal stability up to 300 °C. The high value of the ion transference number suggests that the present blend polymer electrolyte is totally ionic in nature. The improved conductivity, high ionic transference number, and broad voltage stability window support the use of prepared polymer electrolyte with long-term cycle stability and thermal stability for the solid state sodium ion batteries.

**Acknowledgment**

One of the authors (AA) is thankful to the Central University of Punjab for providing the fellowship.